\documentclass[aip,reprint,groupedaddress]{revtex4-1}

\usepackage{graphicx}

\begin{document}


\title{Quantum pumping in graphene with a perpendicular magnetic field}


\author{Rakesh P. Tiwari and M. Blaauboer}
\affiliation{Kavli Institute of Nanoscience, Delft University of Technology, Lorentzweg 1, 2628 CJ Delft, The Netherlands}


\date{\today}

\begin{abstract}
We consider quantum pumping of Dirac fermions in a monolayer of graphene in the presence of a perpendicular magnetic field in the central pumping region.
The two external pump parameters are electrical voltages applied to the graphene sheet on either side of the pumping region.
We analyze this pump within scattering matrix formalism and calculate both pumped charge and spin currents. The predicted charge currents are
of the order of 1000 nA, which is readily observable using current technology.

\end{abstract}




\maketitle


By periodically changing the confining potential of a quantum system, quantum pumping allows a net dc current of carriers to flow in the absence of an 
applied bias \cite{but1,spivak,brouwer}. The possibility of generating electrical currents in solid-state nanostructures using this phenomenon has 
attracted considerable attention \cite{altshuler}. The same phenomenon allows us to adiabatically generate spin-polarized currents in solid-state 
nanostructures \cite{mucciolo}, which is of great interest in the field of spintronics.  
Usually the pumped currents are small ($\sim 100$ pA) and difficult to identify, as such a small signal may be obscured by rectification. 
Indeed, quantum pumps realized so far are commonly believed to have demonstrated rectified currents \cite{swites,watson,brouwer2}.  
In this article we propose a quantum pump realized in graphene in the quantum Hall regime. 
We predict large pumped currents ($\sim 1000$ nA) at modest values of gate voltages and 
magnetic fields, which increases the hope for unambiguous demonstration of quantum pumping.

Ever since the synthesis of high-quality graphene \cite{novoselov}, there has been tremendous interest in the properties of this single-layer form of 
carbon \cite{neto}. Partly as a result of its electronic structure \cite{novoselov,zhang,berger,geim}, graphene exhibits many
unusual electronic and magnetic transport properties, such as a unique type of quantum Hall effect \cite{novoselev1,purewal}, ballistic conduction by 
massless Dirac fermions \cite{novoselev1,purewal}, a size-dependent band gap \cite{han} and large magnetoresistance \cite{hill,geim,cho,ts1}.
Recently some quantum pumps based on graphene were proposed. Zhu and Chen \cite{Zhu} considered a monolayer charge pump with two gate voltages as 
pumping parameters.
Prada \textit{et al.} \cite{prada} considered a similar charge pump and found that the current is pumped through evanescent modes in the pumping region. 
In our proposed pump, both charge and spin currents are pumped through traveling modes and the magnitudes are much larger ($\sim 1000$ nA), 
making it more desirable for applications.  

\begin{figure}
\begin{center}
\includegraphics[scale=0.6,angle=0]{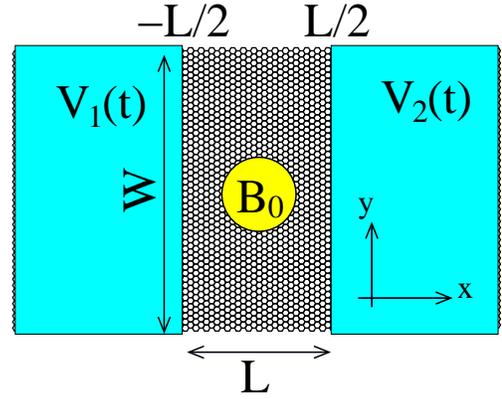}
\caption{(color online). Schematic of the proposed quantum pump. An external field $B_0\hat{z}$ is applied in the central pumping region. The gate
voltages $V_1(t)$ and $V_2(t)$ are the two pumping parameters.}
\label{fig:Fig1}
\end{center}
\end{figure}

Fig. \ref{fig:Fig1} shows a schematic of the device we consider. On the left and right end of the graphene sheet two external voltages 
$V_1(t)=V_{1}+\delta V_1 \sin (\omega t)$ and $V_2(t)=V_{2}+\delta V_2 \sin (\omega t + \phi)$ are applied. An external field $B_0\hat{z}$ (perpendicular 
to the graphene sheet) is applied in the central part of the sheet of length $L$, which forms the pumping region.
We consider short and wide pumps (width($W$) $\gg$ Length($L$)) for which the microscopic 
details of the edges become insignificant and the pumped current is obtained by integrating over the transverse modes.
Also, we assume that the dimensions of the pump are much larger than the carbon-carbon lattice constant. In that case the intervalley tunneling is 
suppressed and it is sufficient to consider just one valley \cite{ando}.

We use the scattering matrix formalism for calculating the pumped current \cite{brouwer}. We divide the system into three regions. In region 1
({$x < -L/2$}) and region 3 ($x > L/2$) gate voltages $V_1(t)$ and $V_2(t)$ are applied, respectively. 
In these regions both up spin and down spin carriers obey the Dirac
equation of motion described by the Hamiltonian $\mathcal{H}=v_F{\bf {\sigma}}\cdot{\bf p}+eV_j$, where $j=1$ for region 1 and $j=2$ for region 3, 
${\bf p}$ is the momentum operator, $v_F$ is the Fermi velocity and ${\bf {\sigma}}=(\sigma_x,\sigma_y,\sigma_z)$ represents the three Pauli matrices. 
In region 2, we describe the effect of the magnetic field $B_0$ and include the effect of Zeeman splitting. We choose the Landau gauge, such that the vector 
potential is ${\bf A}(x)=(0, B_0 x,0)$. Then in region 2 the Hamiltonian for up spin and down spin carriers is 
$\mathcal{H}_{\uparrow}=v_F{\bf \sigma}\cdot({\bf p}+e {\bf A}(x))+g^{\ast}\mu_B B_0/2$ and 
$\mathcal{H}_{\downarrow}=v_F{\bf \sigma}\cdot({\bf p}+e {\bf A}(x))-g^{\ast}\mu_B B_0/2$ respectively. Here $g^{\ast}$ is the spin g-factor 
(assumed to be 2) \cite{gfactor} of the carriers and $\mu_B$ represents the Bohr magneton. 
To simplify the notation we introduce dimensionless units: $l_{B}=\sqrt{\hbar/(eB_0)}$, ${\bf r}\rightarrow l_B {\bf r}$, 
$q \rightarrow q/l_B$, $k_x\rightarrow k_x/l_B$, $E\rightarrow E_0 E$, $E_0=\hbar v_F/l_B$. $\hbar k_x$ represents the momentum along the \textit{x}-axis.

In the adiabatic regime and for (bi)linear response the spin-dependent pumped current through each mode into the left lead is given by \cite{brouwer}: 
\begin{equation}
I_p^{\sigma} = \frac{\omega e \sin{\phi} \delta V_1 \delta V_2}{2\pi} {\rm Im}\left[\frac{\partial r^{\sigma\ast}}{\partial V_1}\frac{\partial r^{\sigma}}{\partial V_2}
+ \frac{\partial t^{\sigma\ast}}{\partial V_1}\frac{\partial t^{\sigma}}{\partial V_2}\right].
\end{equation}
Here $r^{\sigma}$ and $t^{\sigma}$, $\sigma=\uparrow$ or $\downarrow$, represent the coefficients for reflection and transmission os carriers with spin $\sigma$ into the left reservoir 
(a similar expression can be derived for pumping into the right reservoir). The pumped charge current is $I_{\rm p}^{\rm charge}=I_{p}^{\uparrow} + I_{p}^{\downarrow}$ and spin current is $I_{\rm p}^{\rm spin}=I_{p}^{\uparrow} 
- I_{p}^{\downarrow}$. To calculate $r^{\uparrow}$ and $t^{\uparrow}$, we write down the wavefunctions in the three regions corresponding to the respective 
Hamiltonians, and then demand that the wavefunction is continuous across the boundaries at $x=-L/2$ and $x=L/2$. Due to the translational invariance of the geometry in the transverse direction, the momentum along the $y$-axis is conserved and is denoted as $\hbar q$.   
In region 1 the wavefunction is given by:
\begin{equation}
\Psi_1=e^{i q y} \left(\begin{array}{c}
e^{i k_x (x+L/2)} + r^{\uparrow} e^{-i k_x (x+L/2)} \\
e^{i k_x (x+L/2)}e^{i\phi_1} + r^{\uparrow} e^{-i k_x (x+L/2)}e^{{-i\phi_1}}
\end{array}
\right),
\end{equation}

where $\tan\phi_1 \equiv q/k_x$ and the energy of these excitations is given by $E=(eV_1+\hbar v_F\sqrt{k_x^2+q^2})/E_0$. 
Similarly the wavefunction in region 3 is given by:
\begin{equation}
\Psi_3=t^{\uparrow}e^{i q y} \left(\begin{array}{c}
e^{i k_x^{\prime} (x-L/2)}  \\
e^{i k_x^{\prime} (x-L/2)}e^{i\phi_2} 
\end{array}
\right),
\end{equation}
where $\tan\phi_2 \equiv q/k_x^{\prime}$ and the energy of these excitations is given by $E=(eV_2+\hbar v_F\sqrt{k_x^{\prime 2}+q^2})/E_0$. 
In the central region the wavefunction can be written as a linear combination of Weber functions \cite{masir},
\begin{equation}
\Psi_2=e^{i q y} \left(\begin{array}{c}
C_1 D_{p-1}(z) + C_2 D_{p-1}(-z)  \\
\frac{i\sqrt{2}}{E}[C_1 D_{p}(z) - C_2 D_{p}(-z)]
\end{array}
\right),
\end{equation}
where $z=\sqrt{2}(x+q)$ and $p=(E-\frac{g^{\ast}\mu_B B_0}{2E_0})^2/2$. Setting $\alpha_3=D_{p-1}(\sqrt{2}(q-L/2))$, $\alpha_4=D_{p-1}(\sqrt{2}
(-q+L/2))$, $\gamma_1=\frac{\sqrt{2}}{E}D_{p}(\sqrt{2}(q-L/2))$, $\gamma_2=\frac{\sqrt{2}}{E}D_{p}(\sqrt{2}(-q+L/2))$, 
$\omega_1=D_{p-1}(\sqrt{2}(q+L/2))$, $\omega_2=D_{p-1}(\sqrt{2}(-q-L/2))$, $\eta_1=\frac{\sqrt{2}}{E}D_{p}(\sqrt{2}(q+L/2))$ and $\eta_2=
\frac{\sqrt{2}}{E}D_{p}(\sqrt{2}(-q-L/2))$, we find

\begin{eqnarray}
t^{\uparrow}&=&\frac{2i(\eta_2\omega_1+\eta_1\omega_2)\cos{\phi_1}}{f^{+}g^{+}-f^{-}g^{-}} \\
r^{\uparrow}&=&\frac{f^{-}h^{+}-f^{+}h^{-}}{f^{+}g^{+}-f^{-}g^{-}},
\end{eqnarray}

where $f^{+}=i\eta_2+\omega_2e^{i\phi_2}$, $f^{-}=i\eta_1-\omega_1e^{i\phi_2}$, $g^{+}=i\gamma_1+\alpha_3e^{-i\phi_1}$, 
$g^{-}=i\gamma_2-\alpha_4e^{-i\phi_1}$, $h^{+}=i\gamma_2+\alpha_4e^{i\phi_1}$ and $h^{-}=i\gamma_1-\alpha_3e^{i\phi_1}$. Similar expressions are found 
for $r^{\downarrow}$ and $t^{\downarrow}$, the only difference being that $p=(E+\frac{g^{\ast}\mu_B B_0}{2E_0})^2/2$ instead of 
$p=(E-\frac{g^{\ast}\mu_B B_0}{2E_0})^2/2$.

Now we present the numerical results for the pumped charge and spin currents obtained using the above expressions for the reflection and 
transmission coefficients. The current is calculated using Eq. (1) and we show the total pumped currents which are obtained by integrating over all the modes, i.e. $I_{\rm p,T}^{\rm charge}=\int_{-\pi/2}^{\pi/2}\cos({\phi_1})I_{\rm p}^{\rm charge}d\phi_1$ and $I_{\rm p,T}^{\rm spin}=\int_{-\pi/2}^{\pi/2}\cos({\phi_1})I_{\rm p}^{\rm spin}d\phi_1$.  
We choose $W=5$ $\mu$m, $L=0.5$ $\mu$m, $V_{1}=V_{2}=0.1$ V, $\delta V_1=\delta V_2 =0.1$ mV, $v_F=10^6$ m/s and $\omega/(2\pi)=5$ GHz. 
\cite{Zhu,gfactor}. The phase difference between the two external voltages $\phi$ is chosen to be $\pi/2$ so as to maximize the pumped current.
First we calculate the pumped current for fixed energy $E=\mu=100.1$ meV. $\mu$ represents the Fermi energy in the two leads and the 
single 
particle state energy in the central pumping region. 
All these energies are taken to be equal in order to eliminate the possibility of generating current by applied external bias and secure 
energy-conserved tunneling. The temperature is set to zero in all our calculations and we ignore electron relaxation processes. $\mu$ is chosen to be 100.1 meV  
in order to include the contribution of the lowest lying excitations for $V_{1}=V_{2}=0.1$ V and $\delta V_1=\delta V_2 =0.1$ mV. 

\begin{figure}
\begin{center}
\includegraphics[scale=0.33,angle=270]{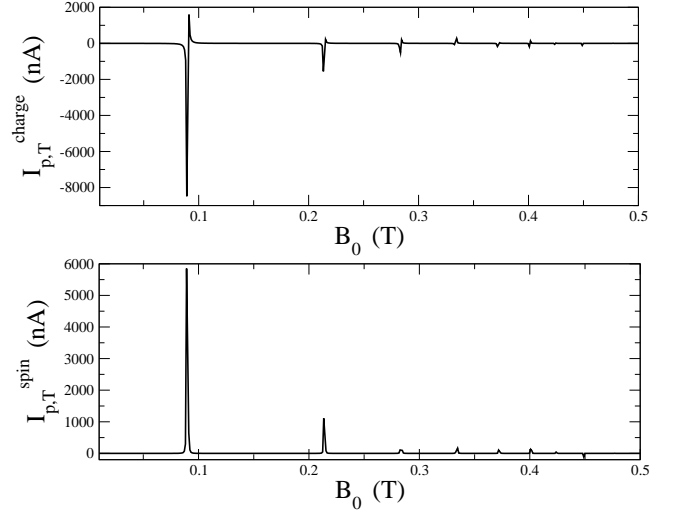}
\caption{Upper panel: pumped charge current as a function of $B_0$. Lower panel: pumped spin current as a function of $B_0$.}
\label{fig:Fig4}
\end{center}
\end{figure}

The upper panel of Fig. 2 shows the pumped charge current as we change the magnetic field for the above mentioned parameters. The magnetic field is varied from 0.01 T to 0.5 T. As expected, the pumped currents are very sensitive to the magnetic field. 
The pumped current is large 
when the magnetic length $l_B$ is comparable to the dimensions of the pump. For $B_0 = 0.1$ T, $l_B \sim 100$ nm and we see large values of the pumped 
currents ($\sim$ 1000 nA). As we increase the magnetic field, the magnetic length decreases and also the total pumped current decreases.
The lower panel of Fig. 2 shows the calculated spin 
current for the same parameters. In general the pumped spin 
current is smaller than the pumped charge current. We believe that the enhancement of the pumped current in the
proposed pump, compared to previously studied pumps without
magnetic field \cite{Zhu,prada} is due to interference arising from the
magnetic length and the dimensions of the pump.
\begin{figure}
\begin{center}
\includegraphics[scale=0.25,angle=270]{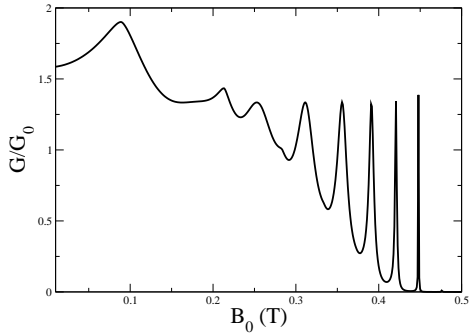}
\caption{ Conductance $G/G_0$ as a function of $B_0$. }
\label{fig:Fig3}
\end{center}
\end{figure}

To compare the calculated pumped currents with rectified currents we calculate the conductance as a function of magnetic field for the 
same parameters (see Fig. 3). The conductance for up-spin carriers is given by $G=G_0\int_{-\pi/2}^{\pi/2}\cos({\phi_1})t^{\uparrow \ast}t^{\uparrow} d\phi_1$ with $G_0=2e^2k_FW/(\pi h)$, where $k_F$ is the Fermi wave-vector. The conductance for 
down-spin carriers (not shown) is almost identical. Peaks in the conductance correspond to a large jump in pumped currents, and the 
pumped charge current 
changes sign around that peak.
This can be used to distinguish between any rectification current and the pumped current. 

Finally we comment on the experimental realization of this pump. Fabricating metallic gates on graphene nanostructures with widths ranging from 
10-100 nm and lengths of 1-2 $\mu$m can be achieved using current technologies \cite{zhang,han}. Perhaps the most difficult part will be to apply 
local magnetic fields of strengths up to 0.5 T, in the central pumping region although efforts have already been made in this direction, for example using 
nanomagnets \cite{nanomag}. The latter can in principle be embedded under the central pumping region of the graphene pump. 
It should be noted that Eq. (1) represents the current pumped throughout one whole pumping cycle with no external bias applied. 
In practice because of the phase 
difference $\phi$ between $V_1(t)$ and $V_2(t)$ (chosen to be $\pi/2$), there will be a small bias across the pump. The pump cycle can be chosen in
a symmetric way such that the bias reverses after half of the period and any rectification current over the whole cycle is canceled.

To summarize, we have proposed a quantum pump based on graphene in a perpendicular magnetic field. 
The pumped current is carried by traveling modes and 
has a large magnitude ($\sim$ 1000 nA). We find that the pumped current has a characteristic dependence on $B_0$. 
Experimental verification of the pump properties would provide a much needed demonstration of the phenomenon of quantum pumping.

This research was supported by the Dutch Science Foundation NWO/FOM.

\end{document}